\documentclass[10pt,conference]{IEEEtran}
\usepackage[numbers]{natbib}
\usepackage{hyperref}
\usepackage{tikz}
\usepackage{graphicx}

\hypersetup{
  pdftitle={Mobility Changes in Response to COVID-19},
  pdfauthor={Warren & Skillman},
  colorlinks=false
}

\newcommand{\round}[1]{\ensuremath{\left\lfloor#1\right\rceil}}

\begin{document}

\title{Mobility Changes in Response to COVID-19}

\author{\IEEEauthorblockN{Michael S. Warren,
    Samuel W. Skillman \\[.3em]}
  \IEEEauthorblockA{Descartes Labs \\
    100 North Guadalupe Street \\
    Santa Fe, NM 87501, USA}\\
  \normalsize{\texttt{v1.0-4-ga6898c6
}}
}

\maketitle

\begin{abstract}

  In response to the COVID-19 pandemic, both voluntary changes in
  behavior and administrative restrictions on human interactions have
  occurred. These actions are intended to reduce the transmission rate
  of the severe acute respiratory syndrome coronavirus 2 (SARS-CoV-2).
  We use anonymized and/or de-identified mobile device locations to measure mobility, a
  statistic representing the distance a typical member of a given
  population moves in a day.  Results indicate that a large reduction
  in mobility has taken place, both in the United States and globally.  In the
  US, large mobility reductions have been detected
  associated with the onset of the COVID-19 threat and specific
  government directives.  Mobility data at the US admin1 (state) and
  admin2 (county) level have been made freely available under a
  Creative Commons Attribution (CC BY 4.0) license via the GitHub
  repository {\tt \href{https://github.com/descarteslabs/DL-COVID-19}{github.com/descarteslabs/DL-COVID-19}}.

\end{abstract}

\vspace{1em}
\noindent

\section{Introduction}

As of March 28, 2020, the coronavirus pandemic COVID-19 is affecting 200
countries and over 600,000
individuals~\cite{zhu_novel_2020,world_health_organization_novel_2020}.
Efforts to prevent spread of the virus include travel and work
restrictions, quarantines, curfews, cancellations and postponements of
events, and facility closures.  These policies aim to reduce the
probability of contact between infected and non-infected persons to
minimize disease transmission~\cite{kissler_social_2020}.  We
hypothesize that the effect of these interventions are measurable in
the aggregate geospatial statistics of people's daily movement.
Measuring changes in mobility is a critical input for forecasting
disease spread and assessing the effectiveness of containment
strategies~\cite{liu_covid-19_2020}.

Geolocation reports from smartphones and other mobile devices offer a
mechanism to sample the movement of individuals.  There are
significant privacy concerns related to the availability of such
data~\cite{warzel_opinion_2019}.  Standard practice
is to make such traces anonymous, where the true identities of nodes
are replaced by random identifiers. However, the privacy concern
remains. Nodes are open to observations in public spaces, and they may
voluntarily or inadvertently disclose partial knowledge of their
whereabouts~\cite{ma_privacy_2010}.  Some studies have shown that sharing
anonymized location data can lead to privacy risks and that,
at a minimum, the data needs to be coarse in either the time domain or
the space domain~\cite{zang_anonymization_2011}.  We do not and will
not use mobile device location data to identify individuals.  All
analysis we perform is statistically aggregated, removing the ability
to characterize the behavior of any single device.

\section{Method}

Cloud computing with associated high-bandwidth storage capacity,
combined with recent advances in machine learning, is enabling
understanding of the world at a scale and level of granularity never
before possible~\cite{warren_seeing_2015}.  Mobile devices
exemplified by the smartphone are an obvious result of this
technological advance.  In this work, we analyze a commercially
available mobile device location dataset using cloud computing
resources.  We heavily leverage our festivus virtual file system
layer~\cite{warren_data-intensive_2017} for this analysis, reading
over 50 TB of input data and utilizing about 50,000 CPU hours of
computation.

Raw location data from commercial data providers is generally provided
in a compressed, comma-delimited text format, with one position report
per line.  Reports contain, at a minimum, an anonymized node id, epoch
time (seconds elapsed since 00:00:00 UTC on January 1, 1970),
latitude, longitude and an estimate of position accuracy.  Data is
obtained in multiple files to facilitate parallel processing, and a
typical volume of compressed data is 100 GB per day.  Data is
delivered once per day.

Our analysis algorithm is implemented with the Python programming
language~\cite{van_rossum_python_1995,oliphant_python_2007} in a
parallel task-based execution environment.
The first step in analysis is to eliminate reports with estimated
position accuracy that exceeds some threshold.  We use a threshold of
50 meters.  Median reported accuracy across all position reports is in
the range of 15-20 meters.

In order to make daily statistics more interpretable, the reported
timestamp is converted to local time.  If a time conversion is not
performed, then the statistics in each area would be analyzed in UTC,
and would not correspond to a ``day'' in the local timezone.  For
instance, a day in UTC from 00:00 to 23:59 would correctly correspond
to a local day (midnight to midnight) in the UK, but would be noon to
noon in New Zealand.  This would lead to analysis artifacts (e.g. a
day that combines Sunday afternoon with Monday morning in New Zealand
cannot be consistently compared to midnight Sunday through midnight
Monday in the UK).  Our initial conversion to local time uses an
approximate local solar time with a time zone offset of,
\[ tzhours = \round{longitude \over 15} \]
when processing individual position
reports, but later determines the precise local timezone once for each
device.

To facilitate the computation of statistics for each node, the reports
for each node are collated.  Device reports are typically not
delivered in a form that can be easily processed per node, so we
perform the equivalent of a parallel bucket
sort~\cite{solomonik_highly_2010}.  Our previous work developing
algorithms for computational cosmology demonstrated the utility of
abstracting the most challenging aspect of a parallel algorithm into a
sort~\cite{warren_2HOT_2013}.

\begin{figure}[htb]
  \centering
  \resizebox{\columnwidth}{!}{\includegraphics{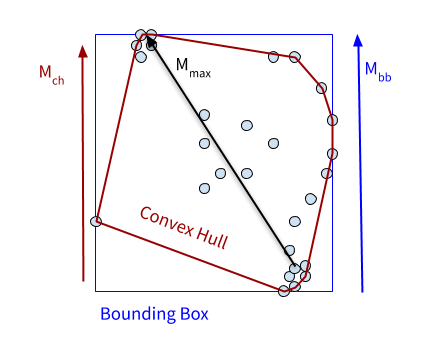}}
  \caption{An illustration of the three mobility measures we calculate
    from a set of positions. $M_{max}$ is the maximum distance to a
    point from the initial point of the day. $M_{bb}$ is the linear
    measure of the bounding box around the points. $M_{ch}$ is the
    linear measure of the convex hull of points.}
  \label{fig:mobility_M}
\end{figure}

We eliminate nodes that have fewer than 10 reports in a day. We also
eliminate nodes that report over a short part of the day (less than 8 hours) to avoid
biasing the mobility stastistic downward with transient reports.  We
calculate a number of statistics intended to capture the mobility of a
population~ (Figure~\ref{fig:mobility_M}).  The max-distance mobility $M_{max}$\ is the maximum Haversine (great
circle) distance (in units of $km$) from the initial location report of
the day.  Outliers (which can occur with bad position reports caused
by poor GPS fixes or other errors) are trimmed by eliminating the top
10\% of the distribution.  The bounding box mobility is computed by determining the area
of the bounding box of the reports $A_{bb}$, and then converting to an equivalent
linear distance in $km$, $M_{bb} = 111 \sqrt{A_{bb}} \cos(latitude)$.
The convex hull mobility is computed by
determining the area of the convex hull $A_{ch}$\ of the reported positions, and then converting
it to an equivalent linear distance in $km$ via $M_{ch} = 111 \sqrt{A_{ch}} \cos(latitude)$.

For a canonical position of each processed node (such as the first or
last report of the day in local time, or a derived location such as
the centroid of node locations) we reverse geocode the location to a
country and administrative region.  We use ISO 3166-1 alpha-2 codes
for the \texttt{country\_code} and GeoNames feature names and codes for
subcategories.  \texttt{Admin1} is the first-order administrative
division, being a primary administrative division of a country, such
as a state in the United States.  \texttt{Admin2} is a second-order
administrative division, such as a county or borough in the United
States.  \texttt{Name} is a populated place feature, representing a city,
town, village, or other agglomeration of buildings where people live
and work.

For each area of interest found by the reverse geocoding procedure
above, we compute a variety of statistics across the mobility measures
$M_{max}, M_{bb}, M_{ch}$.  These statistics include the mean, median
and quartiles.  For the results reported here, we focus on the median
of the max-distance mobility $M_{max}$, which we designate $m50$.  We
further define a normalized mobility index,
\[
m50\_index = 100 {m50 \over m50_{norm}},
\]
where $m50_{norm}$ is a ``normal'' value of $m50$ in a region, defined as the median
$m50$ in that region during a designated earlier time period.  We use
the median weekday value of $m50$ between the dates of 2020-02-17 and
2020-03-07 as $m50_{norm}$ to investigate COVID-19 related changes in
the US.  Note that $m50$ has dimensions of kilometers, while
$m50\_index$ is dimensionless.  $m50\_index$ is easily transformed to
a percentage change from baseline behavior via
\[
pct\ change = m50\_index - 100.
\]

Output of our analysis is stored in two formats.  The first is
newline-delimited JavaScript Object Notation (NDJSON) (also known as
JSON lines format) which is a convenient format for storing structured
data that may be processed one record at a
time~\cite{hoeger_ndjsonndjson-spec_2020}. It works well with
unix-style text processing tools and shell pipelines, and can be read
into a python dictionary data structure with a simple motif
\begin{verbatim}
    for line in file:
        d = json.loads(line)
\end{verbatim}
Using a format such as the
GeoJSON FeatureCollection object is problematic for very large
datasets, because the file must be parsed as a whole, and records
cannot be processed one-at-a-time in a streaming fashion.  We also store
the data in a comma-seperated values (CSV) format which may be more
familiar to users of traditional geographic information systems or
mapping tools.

One significant caveat to the accuracy of the statistics we present
here is related to systematic sampling errors.  We assume that the
sampling of devices is ``fair'' and represents the behavior of the
population as a whole.  Since data is available for only a small
fraction of the total number of devices (a few percent at most), it is
possible that those position reports that are selected for inclusion
in the raw location data we use as input have a correlation with
behavior that is not representative of the average.  For instance, if
the data vendor includes position reports from a transportation
``app'', we expect those reports to show considerably more distance
traveled than an app that would be used while at home.  As long as the
data from both sets of apps is consistently included, the systematic
sampling error would be expected to be small.  However, if the data
vendor loses access to one type of app, that would skew the mobility
statistics, and not represent a change in actual behavior.
Similar types of error would be expected if there was a correlation
between changes in mobility, and the probability of appearing in the
location dataset (e.g. if people who travel long distances are more
likely to disable location reporting).  We mitigate the effects of
these sampling errors to the extent possible by analysis of multiple
datasets, and we seek to validate the statistics through independent
observations.



\section{Results}

On January 23, 2020 the city of Wuhan was closed by canceling planes
and trains leaving the city, and suspending buses, subways and ferries
within it. By this
date at least 17 people had died and more than 570 others had been
infected, including in Taiwan, Japan, Thailand, South Korea and the
United States.  Ideally, we could analyze mobility around the
epicenter of the outbreak in China, but we do not have access to data
from that country.  A recent study from researchers in Beijing was able
to access aggregated statistics provided by the two largest
telecommunications operators in China~\cite{cao_incorporating_2020}.

\begin{figure}[htb]
  \centering
  \resizebox{\columnwidth}{!}{\includegraphics{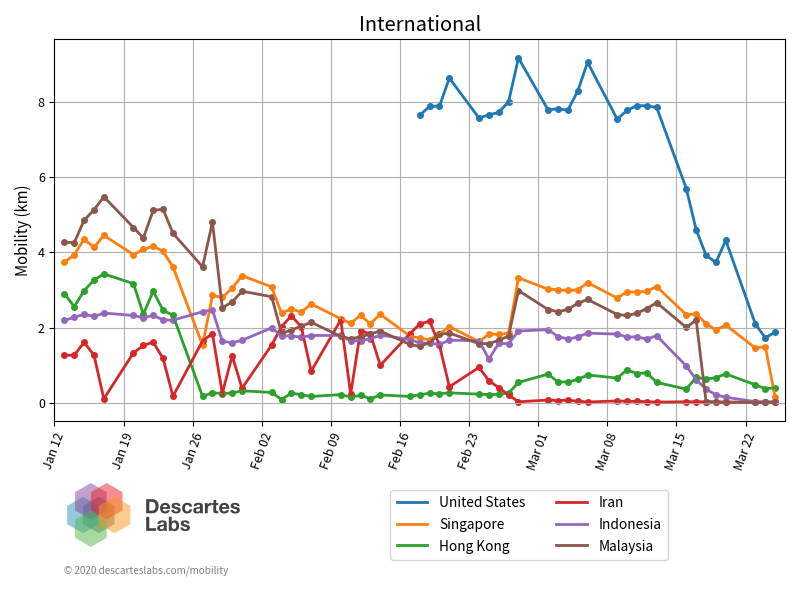}}
  \caption{Change in mobility internationally since mid-January for
    select countries.  Between January 24 and 27, there is a dramatic
    drop in mobility observed in Hong Kong and Singapore corresponding
    to the time the spread of COVID-19 was recognized.  See also the
    next figure for an alternate view of the same countries.}
  \label{fig:int}
\end{figure}

Between January 24 and 27, there is a dramatic drop in
mobility observed in Hong Kong and Singapore~\ref{fig:int}.  On
Friday, Jan.~24 $m50$ in Singapore was 3.6 km.  By Monday, Jan.~27 it
had fallen more than 50\% to 1.5 km.  The change was even more
pronounced in Hong Kong, dropping to 180 meters from 2.3 km.  The
weekend of January 25 also corresponded to the Chinese New Year
holiday, so the change in mobility in that time period could have been
independently influenced by the holiday.

\begin{figure}[htb]
  \centering
  \resizebox{\columnwidth}{!}{\includegraphics{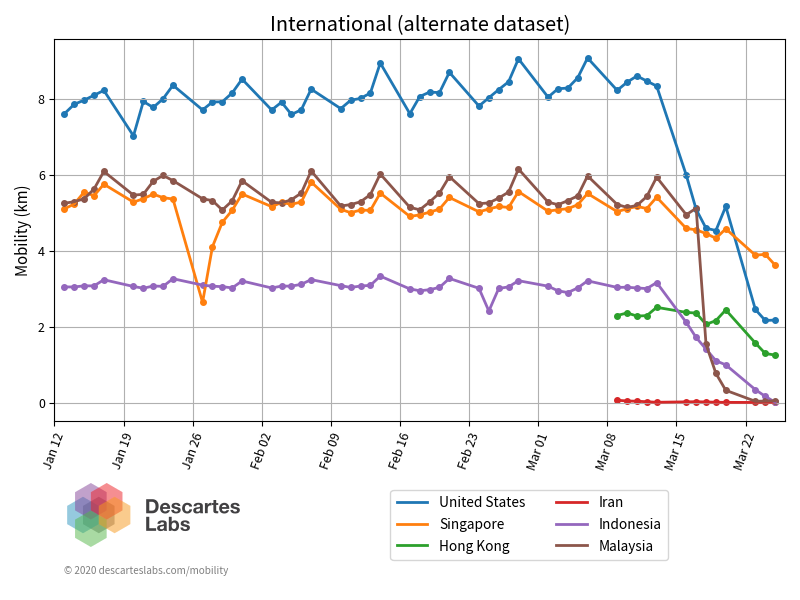}}
  \caption{Change in mobility internationally since mid-January for
    select countries, using an alternate dataset.  The drop and
    recovery in Singapore on Jan.~27 is more clearly represented.
    Peaks on Friday are seen in the US and Malaysia, but less so in
    Indonesia. The drop on Feb.~25 in Indonesia corresponds to a day of heavy rain and flooding.}
  \label{fig:int2}
\end{figure}

The behavior of the these graphs for some countries between January 29
and February 27 is subject to the systematic sampling error we
described previously, which shows an anomalous drop in mobility during
the month of February.  Using an alternate input dataset, the behavior
in Singapore and Malaysia is much more consistent over that time
period~\ref{fig:int2}.  Other notable features from that graph are the
weekly cycles, with Friday having the highest mobility, particularly
in the US, a lower value on Jan.~20 (the Martin Luther King Jr. Day
holiday in the US) and the dip on Feb.~25 in Indonesia, which
corresponds to a day of heavy rain which flooded Jakarta, the capital.
Figures~\ref{fig:int}-\ref{fig:int2} also show the significant response
in the US beginning March 14, and corresponding drops in the rest of the
world as a new wave of mobility restrictions were put into place.
Note that we do not plot values on Saturday and Sunday for clarity, since
mobility on Sunday is usually much lower than weekdays.

\begin{figure}[htb]
  \centering
  \resizebox{\columnwidth}{!}{\includegraphics{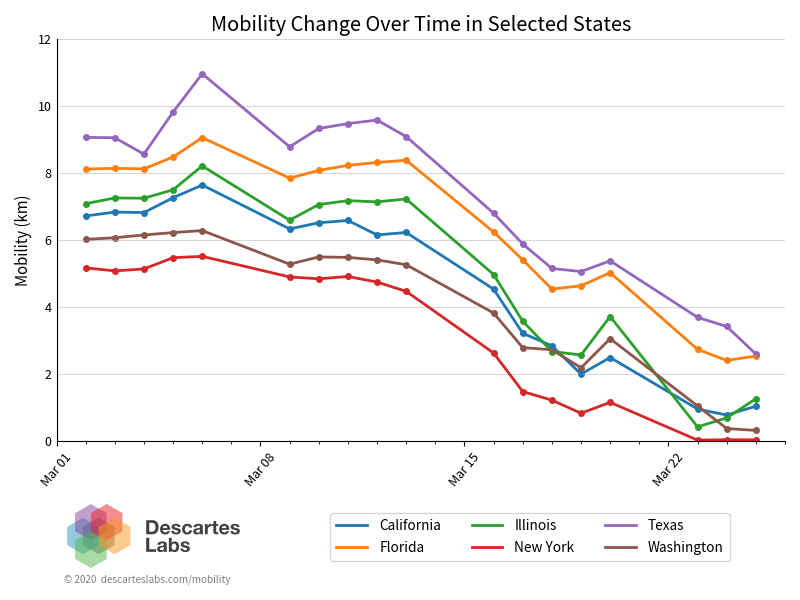}}
  \caption{Change in mobility over the first three weeks of March, 2020
    in selected states in the US. Drops in mobility are detected
    across all six states March 13-16, with New York falling from 5.2
    km on Monday, March 2 to only 31 meters on Monday, March 23.  This
    implies that half of the individuals in the entire state of New
    York spent most of their day within less than 100 feet of their
    initial position.}
  \label{fig:us6km}
\end{figure}

\begin{figure}[htb]
  \centering
  \resizebox{\columnwidth}{!}{\includegraphics{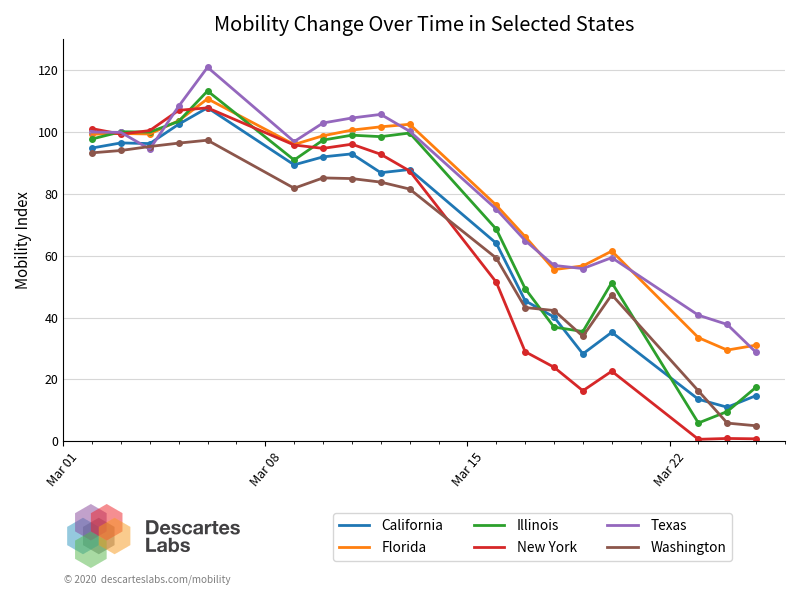}}
  \caption{Change in mobility index (normalized to 100 for the weeks
    of Feb.~17 through Mar.~7 2020). Florida and Texas
    have dropped to about 30\% of normal, and California, Illinois,
    New York and Washington to less than 20\% of normal.}
  \label{fig:us6}
\end{figure}

In Figure~\ref{fig:us6km} we show mobility in six selected US states
beginning March 2.  We detect that more rural states (with presumably
  longer travel distances) normally have a higher $m50$ than more
  urban states, specifically values of 9-11 km for Texas vs 5 km for
  New York for the first week in March.  Reductions in mobility are
  detected across all six states March 13-16, with New York falling
  from 5.2 km on Monday, March 2 to only 31 meters on Monday, March
  23.  This implies that half of the individuals in the entire state
  of New York spent most of their day within less than 100 feet of
  their initial position.  To better understand the change from
  ``normal'' mobility and compare states with each other,
  Figure~\ref{fig:us6} plots the normalized $m50\_index$, demonstating
    that Florida and Texas have dropped to about 30\% of normal, and
    California, Illinois, New York and Washington to less than 20\% of
    normal.

\begin{figure}[htb]
  \centering
  \resizebox{\columnwidth}{!}{\includegraphics{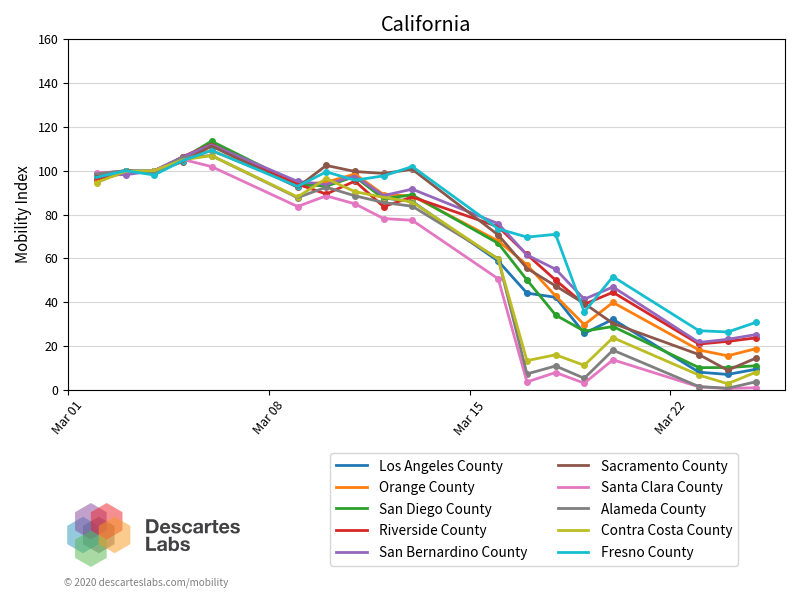}}
  \caption{Change in mobility index for the largest counties in
    California. The inital reduction seen on March 17 in Alameda,
    Contra Costa and Santa Clara counties can be attributed to the
    shelter-in-place order for the six Bay Area counties which took
    effect at 12:01am on that day.  The order expanded statewide the
    evening of March 19.}
  \label{fig:ca}
\end{figure}

We can also look at smaller geographic regions by plotting the change in
mobility within counties.  In Figure~\ref{fig:ca} we show the 10 most highly
sampled counties in California.  The inital reduction seen on Mar.~17 in Alameda,
Contra Costa and Santa Clara counties can be attributed to the shelter-in-place
order for the six Bay Area counties which took effect at 12:01am.
The order expanded statewide the evening of Mar.~19.

\begin{figure}[htb]
  \centering
  \resizebox{\columnwidth}{!}{\includegraphics{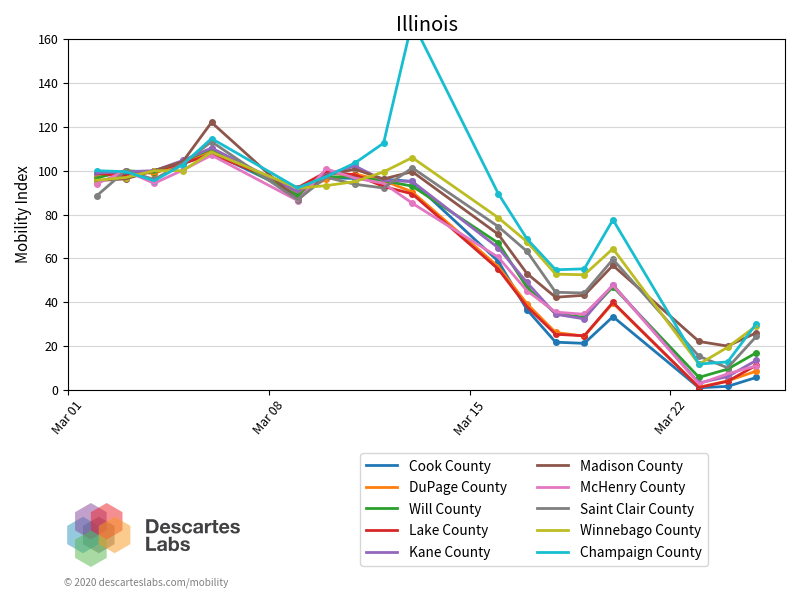}}
  \caption{Change in mobility index for the largest counties in Illinois. The
    spike in mobility on March 13 in Champaign County we believe is
    associated with a large number of students leaving the University
    of Illinois Urbana-Champaign after the on-campus classes were
    cancelled.}
  \label{fig:il}
\end{figure}

\begin{figure}[htb]
  \centering
  \resizebox{\columnwidth}{!}{\includegraphics{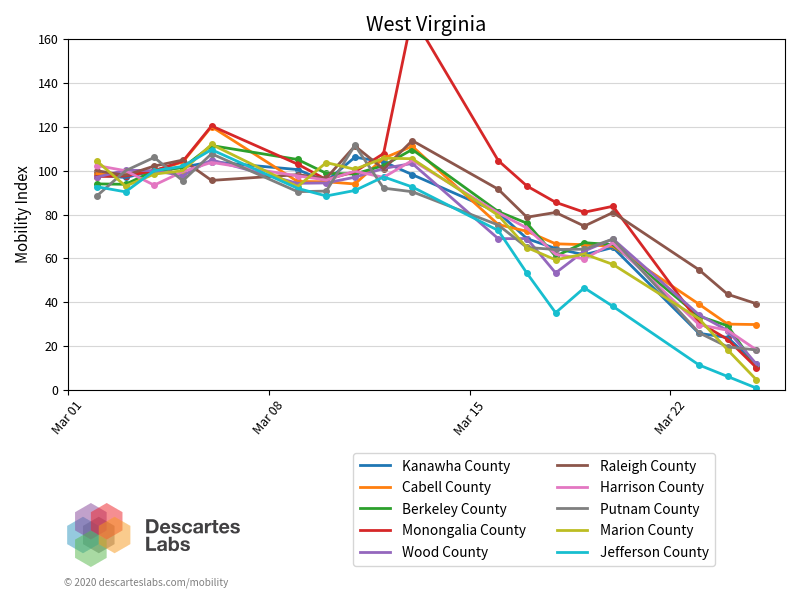}}
  \caption{Change in mobility index for the largest counties in West
    Virginia. The jump on March 13 in Monongalia County is associated
    with students leaving West Virginia University (enrollment of
    30,000 students in a county with about 100,000 people).}
  \label{fig:wv}
\end{figure}

Counties in Illinois are illustrated in Figure~\ref{fig:il}.  The
spike in mobility on March 13 in Champaign County we believe is
associated with a large number of students leaving the University
of Illinois Urbana-Champaign after the on-campus classes were
cancelled.  A similar jump is seen in Figure~\ref{fig:wv} on March
13 in Monongalia County, associated with students leaving West
Virginia University (enrollment of 30,000 students in a county with
about 100,000 people).  West Virgina was the last state in the US
to report a coronovirus infection.





\begin{figure}[htb]
  \centering
  \resizebox{\columnwidth}{!}{\includegraphics{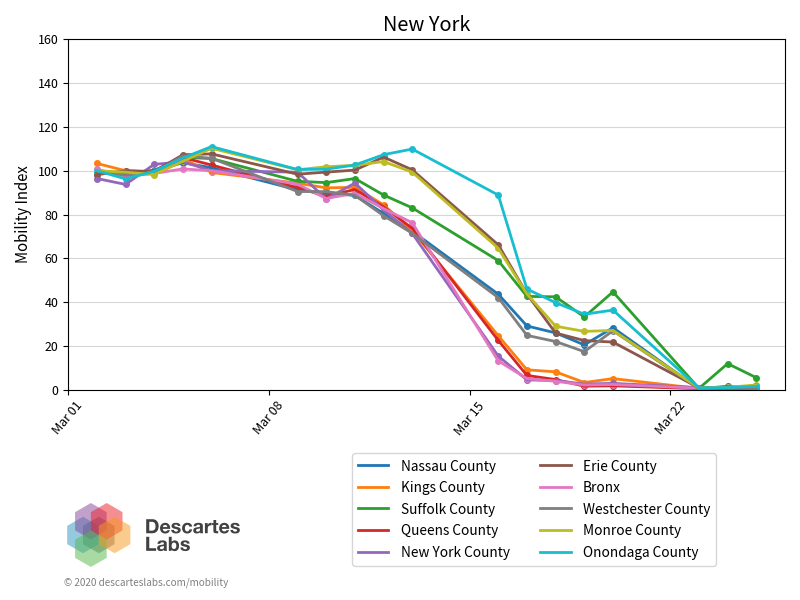}}
  \caption{Change in mobility index for the largest counties in New York.}
  \label{fig:ny}
\end{figure}

\section{Conclusion}

We have detected dramatic changes in mobility due to COVID-19, both
within the US and globally.  Mobility data at the US admin1 (state)
and admin2 (county) level have been made freely available under a
Creative Commons Attribution (CC BY 4.0) license via the GitHub
repository {\tt \href{https://github.com/descarteslabs/DL-COVID-19}{github.com/descarteslabs/DL-COVID-19}}.  As the
situation develops, combining information such as that we derive here
with pandemic growth rates in various geographies will allow more
accurate models of the interventions being made, and help save lives.

\bibliographystyle{myunsrtnat}
\bibliography{mobility}
\end{document}